\documentclass[pdflatex,sn-mathphys-num]{sn-jnl}

\usepackage{graphicx}%
\usepackage{multirow}%
\usepackage{amsmath,amssymb,amsfonts}%
\usepackage{amsthm}%
\usepackage{mathrsfs}%
\usepackage[title]{appendix}%
\usepackage{xcolor}%
\usepackage{textcomp}%
\usepackage{manyfoot}%
\usepackage{booktabs}%
\usepackage{algorithm}%
\usepackage{algorithmicx}%
\usepackage{algpseudocode}%
\usepackage{listings}%
\usepackage{orcidlink}

\theoremstyle{thmstyleone}%
\theoremstyle{thmstyletwo}%

\theoremstyle{thmstylethree}%

\begin{document}

    \title[Article Title]{ProtBoost: protein function prediction with Py-Boost and Graph Neural Networks - CAFA5 top2 solution 
    }


\author*[1,2,3,4]{\fnm{Alexander} \sur{Chervov\orcidlink{0000-0003-4564-3272}}}\email{alexander.chervov@curie.fr, al.chervov@gmail.com}
\equalcont{These authors contributed equally to this work.}

\author[5]{\fnm{Anton} \sur{Vakhrushev}}\email{btbpanda@gmail.com}
\equalcont{These authors contributed equally to this work.}

\author[5]{\fnm{Sergei} \sur{Fironov}}\email{ifserge@gmail.com}

\author[1,2,3,4]{\fnm{Loredana} \sur{Martignetti\orcidlink{0000-0003-1535-7515}}}\email{loredana.martignetti@curie.fr}

\affil[1]{\orgdiv{Institut Curie}, \orgaddress{\city{Paris}, \country{France}}}

\affil[2]{\orgdiv{INSERM}, \orgname{U900}, \orgaddress{\city{Paris}, \country{France}}}

\affil[3]{\orgdiv{Mines ParisTech}, \orgaddress{\city{Paris}, \country{France} } }

\affil[4]{\orgdiv{PSL University},
\orgaddress{\city{Paris}, \country{France} } }


\affil[5]{\orgdiv{Independent Researcher} }

\abstract{Predicting protein properties, functions and localizations
are important tasks in bioinformatics. Recent progress in 
machine learning offers an opportunities for improving existing methods.
We developed a new approach called ProtBoost, which relies on the strength of pretrained protein language models, the new Py-Boost gradient boosting method and Graph Neural Networks (GCN).
The ProtBoost method was ranked second best model in the recent Critical Assessment of Functional Annotation (CAFA5) international challenge with more than 1600 participants.
Py-Boost is the first gradient boosting method capable of predicting thousands of targets simultaneously, making it an ideal fit for tasks like the CAFA challange. Our GCN-based approach performs stacking of many individual models and boosts the performance significantly. Notably, it can be applied to any task where targets are arranged in a hierarchical structure, such as Gene Ontology.
Additionally, we introduced new methods for leveraging the graph structure of targets and present an analysis of protein language models for protein function prediction task. ProtBoost is publicly available at: https://github.com/btbpanda/CAFA5-protein-function-prediction-2nd-place.}


\maketitle

\section{Introduction}\label{sec1}

Understanding protein properties and their functions in living organisms is one of the key challenges in biology. 
Gene Ontology (GO) \cite{ashburner2000gene}
is one fundamental database for the standardized representation of functional annotations across all species. The number of available annotations ("terms") in Gene Ontology is already more than 40,000 and growing each year. Public databases such as UniProtKB \cite{uniprot2023uniprot} 
contain over 245 million protein entries, with fewer than 571,000 of these having been manually annotated. The MGnify database from metagenomic assemblies \cite{uniprot2023uniprot} 
includes 3 billion non-redundant protein sequences, organized into 729 million clusters. Experimentally establishing a link between such a large number of functional annotations and such a large number of proteins is unfeasible. Thus, many computational methods have been developed over the past 20 years to address this task \cite{lee2007predicting} \cite{tiwari2014survey}. However, the function of a large number of proteins remain unknown for many sequenced genomes. The aim of this work is to provide a significant contribution to this fascinating field of research.

One of the key events to benchmark various methods in this field is the CAFA (Critical Assessment of Function Annotation) challenge 
\cite{radivojac2013large} \cite{jiang2016expanded} \cite{zhou2019cafa}.
Organized similarly to the Critical Assessment of protein Structure Prediction (CASP) challenges, participants are tasked in a first stage with predicting protein functions that are not yet known experimentally. The second stage involves experimentally establishing these functions and comparing them with the predictions. Previous CAFA challenges (CAFA1 - CAFA4) took place between 2010 and 2019. In late 2022, the success and scope of the CAFA challenges caught the attention of \textit{Kaggle}, a popular data science competition platform and online community for data scientists and machine learning practitioners (https://kaggle.com). As a result, the fifth edition of CAFA was hosted on the Kaggle platform in 2023. CAFA5 achieved significant success, attracting 1,987 competitors organized into more than 1,600 teams, who submitted 2,850 entries. The work presented in this article provides a detailed description of the ProtBoost model submitted by our team (U900) for the CAFA5 competition, in which we secured the second place in the final ranking.

Our model was designed to address the challenge posed by the CAFA5 competition, focusing on the prediction of protein function annotations. We were able to achieve high accuracy and reliability in our predictions, by leveraging (i) cutting-edge techniques of protein language models, (ii) a gradient boosting algorithm called \textit{Py-Boost} that we developed recently specifically for multi-target tasks \cite{vakhrushev2022pyboost},
(iii) a novel scheme to predict targets organized in a hierarchical tree structure, such as Gene Ontology, that consist in stacking with graph neural networks and (iv) a specific scheme to optimize targets which we call \textit{conditional probability modeling}. 

Previous studies have been based on the assumption that proteins with similar sequences will exhibit similar functions to some extent. Sequence similarity can be measured using classical sequence alignment methods, such as BLASTP \cite{altschul1990basic}. While this idea is fundamentally sound, it encounters certain difficulties, such as determining the appropriate threshold for similarity and difficulties with incorporating other similarities beside sequence. To address these challenges, machine learning approaches offer promising solutions. These approaches can be trained on a subset of proteins with known labels and then validated on unseen subsets. They allow for the integration of various features and the optimization of predictive accuracy, thereby overcoming the limitations of relying solely on sequence similarity. Thus, the selection of thresholds, additional features and model specificity are governed by the quality of the predictions rather than by subjective choices. Consequently, machine learning methods naturally arise as suitable solutions for the task of protein function annotation. 

Let us briefly highlight the ideas behind some recently proposed solutions during the CAFA5 challenge. An interesting strategy based on literature-based information was used in the Top1 \cite{top1CAFA5} and Top4 (ProtGoat \cite{chua2024protgoat}) CAFA5 solutions. This approach encodes textual information from papers mentioning specific proteins into vector of features usable by machine learning models. It poses a risk of overfitting, since newly discovered proteins may lack sufficient literature references. However it proves its effectiveness in CAFA5, presumably because the emergence of literature knowledge frequently precedes the assignment of new GO annotations to proteins. The top-performing model proposed in CAFA5 is called GOCurator \cite{top1CAFA5}, developed by a team led by Prof. Shanfeng Zhu at Fudan University. The same team has participated in the CAFA3 and CAFA4 and achieved the top performance. GOCurator uses a diverse ensemble of approaches, including text mining, protein language models, protein structural information, and GORetrieval — a deep learning based matcher between GO information and protein data. Multiple models are constructed using these approaches and then combined. The diverse range of methods likely played a key role in securing top performances in multiple challenges.

The Top3 \cite{top3CAFA5} CAFA5 solution introduced a novel processing technique for non-experimental annotations available in UniProtKB \cite{uniprot2023uniprot} which correspond to GO labels generated by computational methods without experimental evidence. Instead of using these labels directly as features, this model uses them after an advanced preprocessing via convolutional neural networks (CNN).
The Top5 \cite{top5CAFA5} team (from Freddolino Lab) exploited structural alignment using AlphaFold2, protein-protein interaction networks, and Pfam families into their model. The Top6 \cite{top6CAFA5} team (also from Freddolino Lab) developed a procedure to detect subtle variations in the loss function during neural network training. They implemented a "soft F1 loss" which significantly enhanced model performance compared to the standard binary cross-entropy (BCE) loss. 




Top14 \cite{top14CAFA5} team (Zoltan) shared several remarkable ideas to improve model performance, including downsizing the training dataset to the most annotated proteins and selecting the top frequent targets respecting sub-ontologies. Michele Tinti proposed a scheme \cite{MTmergedatasets} to merge electronic predictions from UniprotKB with models predictions.
The Top19 team (from Hoehndorf Lab \cite{kulmanov2024protein}) developed an innovative approach for working with ontologies. They introduced novel ontology embeddings where each node's embedding is represented as a ball instead of a point. This method allows to translate
ontology axioms into geometric restrictions on the embeddings. And thus enabling zero-shot predictions, i.e. non-trivial predictions for labels which are completely zero in the training set.
Overall, the primary components for most models were pretrained protein language models, which partially adhered to the approach we initially proposed at the start of the challenge. However, to achieve the best results we have implemented multiple and unique strategies in our model which we detail in Section 2.1.

The main contributions of the present work are the following:

\begin{enumerate}
    \item We developed a new method to predict GO-terms associated to proteins, incorporating modern innovative technologies, including protein language models, the Py-Boost library and Graph Neural Networks. 
    \item We demonstrate the efficiency of Py-Boost \cite{vakhrushev2022pyboost}, a new gradient boosting algorithm specifically designed to work with multi-target tasks developed by one member of our team, for protein properties prediction tasks.  
    \item We propose several new techniques designed for predicting variables which belong to hierarchical structure like Gene Ontology. The first one is a conditional probability modeling strategy (CondProbMod) for constructing new models from the existing ones and the second one is a specific modification of Graph Neural Network (GCN) to stack predictions of several models.

\end{enumerate}

All these novelties have been implemented in a new model and package, called ProtBoost, to predict protein functions. This model was trained on a dataset including 142246 proteins representing eukaryotes, prokaryotes and viruses.

\section{Results}


\subsection{Model overview }
\begin{figure}[h]
\centering
\includegraphics[width=0.9\textwidth]{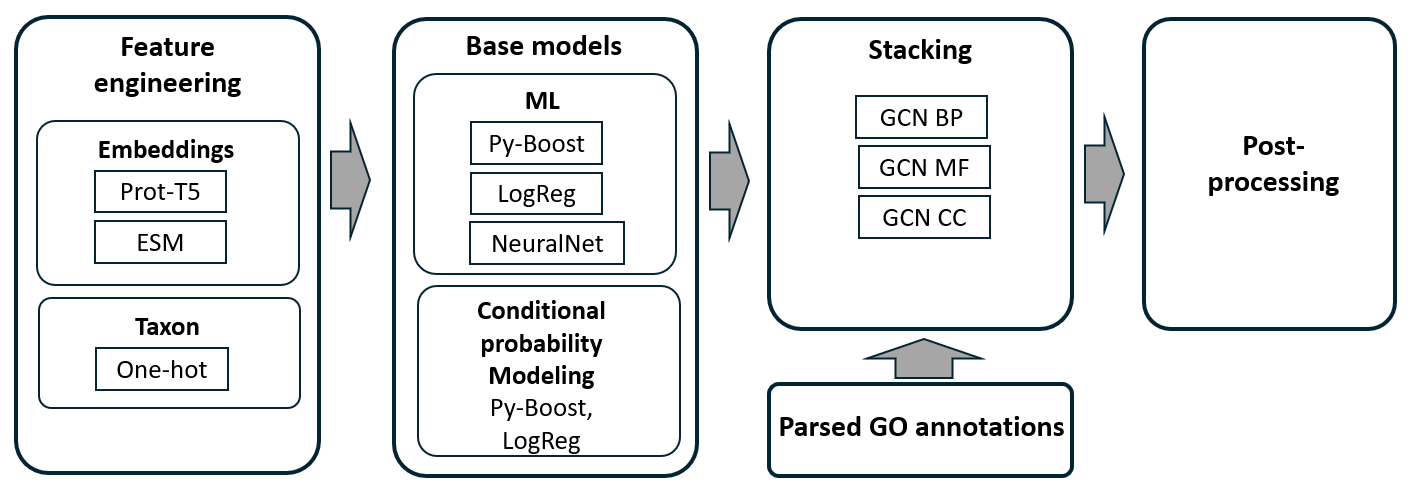}
\caption{Architecture of the ProtBoost modeling scheme. The main blocks are: 1. Feature engineering. 2. Base models and their modified versions by Conditional Probability Modeling approach. 3. Stacking with Graph Neural Network and incorporation of the electronic annotations. 4. Post-processing.  }\label{modelling_overview_CAFA}
\end{figure}

The main architecture of the ProtBoost model is illustrated in Figure \ref{modelling_overview_CAFA}.

The main analytical steps are the following.

\begin{enumerate}
\item \textbf{Feature engineering.} Here, the main features are built with the help of protein language models, including pretrained language models such as Prot-T5 \cite{elnaggar2021prottrans} and ESM2 \cite{lin2022language} that take in input protein sequences and output numerical vectors enabling to capture the underlying information encoded in these sequences. The size of these vectors is the same for all proteins independently on the sequence length. Beside the sequence embeddings, we also included in the model taxon information.

\item \textbf{Base models.} Our core model is Py-Boost, a novel gradient boosting algorithm that we specifically developed for multi-target tasks \cite{vakhrushev2022pyboost}). The model takes protein embeddings along with taxon information as input and predicts GO terms. These terms are encoded in a binary vector indicating whether the terms are associated with the input protein or not. We restrict the prediction only to the most frequent terms, reducing the initial database of 40000 terms to 5000-13000 terms depending on the model. We assign a prediction equal to zero for terms not included into the model. In addition to Py-Boost, we trained a neural network model and a logistic regression model that we expect to better predict only least frequent terms, due to the simplicity of the model and its strong regularization. Another key procedure implemented at this step is what we call conditional probability modeling (CondProbMod), specifically designed to predict targets organized into a hierarchical tree structure like an ontology. CondProbMod allows to run the base models including conditional probabilities for GO term inference that are based on the structure of the ontology (see Methods).   

\item \textbf{Stacking by graph neural network (GCN)}. Stacking in machine learning is a widely used technique where multiple models, also known as base learners or base models, are trained to solve the same problem, and their predictions are combined by a downstream model called a meta-model. Since targets of our models belong to an ontology graph, it is natural to use GCN-like approach for two purposes, combining prediction of different base models and aggregating them across the graph structure. At this step, we integrate additional features to aggregate not only our models predictions, but also other predictions available in literature including QuickGO, UniprotKB and other resources providing electronic annotations of protein function. 

\item \textbf{Post-processing}. 
The last step consists in post-processing the output predictions, based on the conventional assignment of GO labels in the CAFA challenges, that is if a child node have a label "1", then all its parent nodes will be assigned label "1" automatically.
We implemented a procedure to leverage the property 
that the values of parent nodes are always greater than or equal to those of their child nodes.

\end{enumerate}

The models were trained using random 5 fold cross-validation scheme with GroupKFold on unique  sequence (see Methods). The stacking was tested on small holdout subsample.

\subsection{Performance results}

\begin{figure}[h]
\centering
\includegraphics[width=0.9\textwidth]{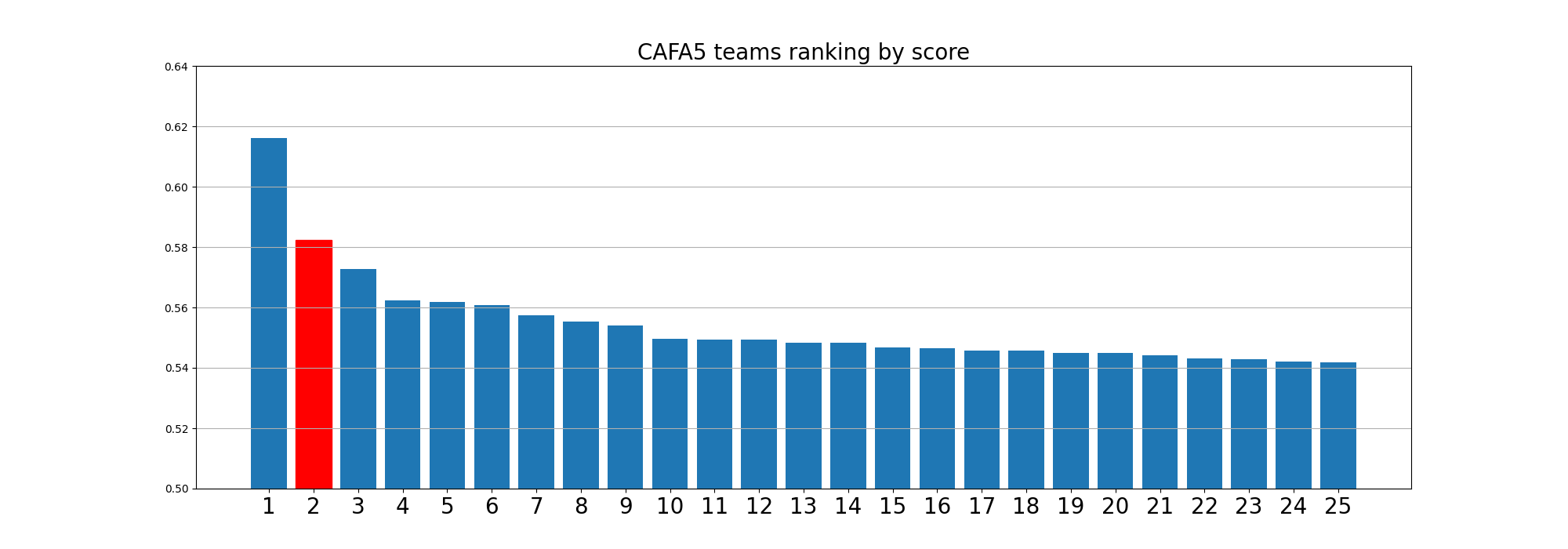}
\caption{Final score of the top teams on CAFA5 by the CAFA-metric (modified F1-score). The score of the ProtBoost model is highlighted in red.}\label{leaderbord_CAFA}
\end{figure}

We trained our model on a comprehensive dataset specifically curated for the task of protein function prediction (see Methods). The dataset included 142246  protein sequences and their corresponding functional annotations borrowed from the the UniProtKB (see subsection  \ref{ss_data}).
At the final stage of the challenge, some proteins whose function were unknown experimentally have received experimental verification. These proteins serve as the benchmark for testing the models, with a final score being assigned based on their performance. Our ProtBoost model achieved the second best performance on the final leaderboard of the CAFA5 challenge, with a score of 0.58240. Our score was significantly higher than those of the third and fourth place models, which scored 0.57276 and 0.56245, respectively. Final results of the top 25 ranked models are reported in Figure \ref{leaderbord_CAFA}.
The metric used in CAFA challenges is a modification of F1-score described in subsection  \ref{ss_eval_m}.

\subsection{Ablation study }


We reported the preliminary scores published by the organizers throughout the various phases of the CAFA5 challenge. The scores obtained during the challenge were recorded after key improvements of the model. This can be considered an ablation study and demonstrates that each component of our model contributes to improving performance. Results have been reported in Table \ref{tab:ablation_study}.

\begin{table}[ht]
\centering
\begin{tabular}{|l|l|l|}
\hline
Step & Added/Changed component & Score  \\
\hline
1 & Baseline: Ridge regression over Prot-T5 protein language model embeddings   & 0.47 \\
\hline
2 & Single Py-Boost over Prot-T5 + conditional probability modelling  & 0.51 \\
\hline
3 & Added: taxon features (one-hot for top frequent taxons)  & 0.53 \\
\hline
4 & Conditional probability modeling for Py-Boost and LogReg (blended  with direct models) & 0.55 \\
\hline
5 & Added: external data - Gene Ontology electronic Annotations (GOA) & 0.57 \\
\hline
6 & Added: Ensemble with Neural Network base model & 0.59 \\
\hline
7 & Stacking all models and GOA by Graph Convolutional Network (GCN) & 0.62 \\
\hline
\end{tabular}
\caption{\label{tab:ablation_study} Scores obtained after adding or modifying key components of the model}
\end{table}

\section{Discussion}\label{sec3}

Here, we present the ProtBoost model developed in the context of the CAFA5 challenge to predict the biological function of proteins. ProtBoost is based on novel machine learning techniques, including protein language models, Py-Boost library, GCN-stacking and conditional probability modelling applied to ontology structured labels. The model was ranked second in the final ranking of the CAFA5 competition.

This study, along with the CAFA5 competition, enables us to draw several conclusions. First, protein language models have become an effective and indispensable tool for studying protein properties and functions. Second, Py-Boost demonstrated its efficiency in this new context, characterized by the use of feature embeddings for proteins and the presence of thousands of binary targets with highly varied distributions. Overall, it has proven to be an innovative tool very effective and easy to use in this context of CAFA like tasks where multiple targets need to be predicted simultaneously. Independent analyses by other studies \cite{luukkonen2023large} confirm Py-Boost superiority over the other boosting for multi-target tasks.
Finally, the novel CondProbMod technique that incorporates the hierarchical structure of GO has proven to be highly useful to solve the problem of protein function assessment. Results of the ablation study show that  each component of our model provided significant improvement of the performance. This validates the significance of our findings, suggesting they can be effectively integrated with other researchers' approaches in future work.

The CAFA5 challenge, organized on the Kaggle platform, attracted a large community of 1,987 participants. These challenge fosters a collaborative research environment where ideas and techniques are shared and tested through public notebooks and discussion posts. Our team greatly benefited from the community's shared knowledge and also made significant contributions, with three of our notebooks ranking among the top ten most voted. Additionally, we organized several webinars, such as the CAFA5 discussion, where CAFA5 organizers and top experts shared their insights.

We believe Kaggle has become an invaluable resource for the research community, not only by hosting challenges but also by providing free cloud access with GPU and TPU capabilities, unlimited storage for publicly available data, and a collaborative workspace for projects.  This platform also hosts a vast repository of code and data resources across various fields of bioinformatics, making it an indispensable tool for researchers. 

With the significant advancements in artificial intelligence methods, especially the great success of AlphaFold \cite{jumper2021highly}, it is expected that similar techniques will have a comparable impact on protein function prediction tasks. However, this level of success has not yet been achieved. One of the key challenges lies in the available data. Collecting, curating, and maintaining extensive databases like Gene Ontology is a complex task, even more challenging than managing three-dimensional protein structure databases. Consequently, the training data for machine learning models in this domain is often neither clean nor complete.
In addition to technical and organizational difficulties, there are conceptual challenges. Experimental verification of protein functions can yield varying results across different studies. For example, comparing several human cell cycle studies reveals only about one hundred common genes, despite each study identifying around a thousand genes \cite{grant2013identification},\cite{giotti2019assembly}. This raises the question of the feasibility and extent of establishing an objective ground truth.
Based on our previous cell cycle studies \cite{chervov2022computational}, we consider the binary nature of labels in Gene Ontology to be one of the primary challenges. The binarization of continuous variables involves arbitrary threshold decisions, which can lead to discrepancies across different biological studies. It might be more biologically accurate to use continuous labels that reflect the degree of involvement of proteins in certain processes. Therefore, biological ground truth exists and is accessible with current technologies, but careful interpretation of experimental results is required.
Overall, we remain optimistic about future progress in modeling protein functions. The advances in both data availability and machine learning algorithms are promising. 

\section{Methods}\label{sec4}

\subsection{Py-Boost}
\begin{figure}[h]
\centering
\includegraphics[width=0.9\textwidth]{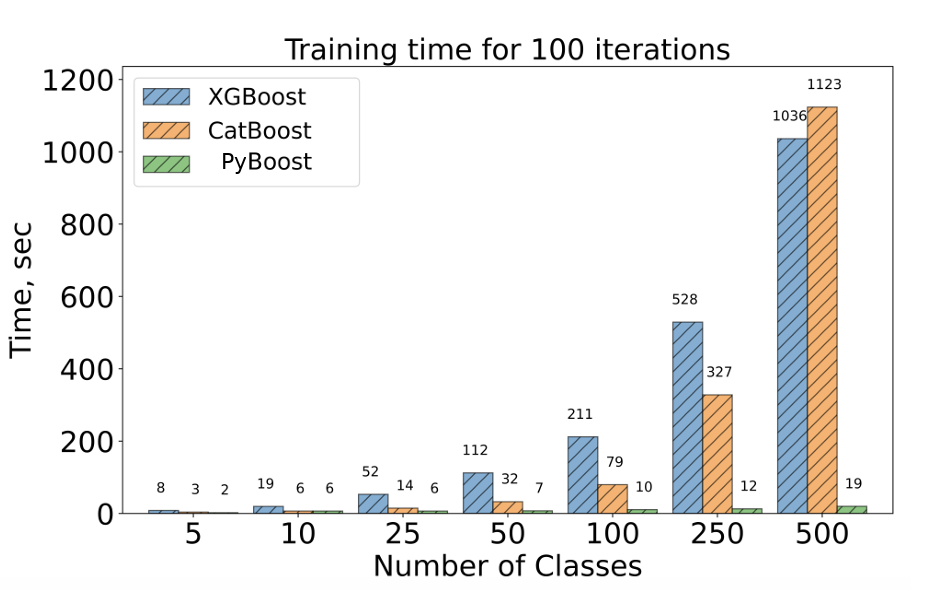}
\caption{Benchmark of Py-Boost against XGBoost and CatBoost (on GPU). With number of targets growing Py-Boost achieves over 50 times speedup vs other gradient boostings. Speedup is achieved due to principal algorithmic innovation in approximating the loss function  in multi-target cases ("sketching") as well as efficient GPU-based implementation. 
During CAFA we trained Py-Boost models 
for 4500 targets coming from different sub-ontologies in the following way: 3000 top frequent BP, 1000  MF,  500 top frequent CC.
}\label{pyboost_histogram2}
\end{figure}

Gradient boosting provides effective and easy to use way to build supervised learning
models for tabular data. XGboost, LightGBM and CatBoost are some of the most widely used packages by practitioners. However they cannot be effectively used for tasks with multiple targets due to prohibitive computational time. Py-Boost is a new gradient boosting algorithm developed specifically for multi-target problems \cite{vakhrushev2022pyboost}. It speeds up the computations significantly (e.g. more than 50 times for the tasks with 500 targets comparing to the other gradient boostings, as shown in Figure \ref{pyboost_histogram2}). The more targets in the task, the greater the speedup achieved by Py-Boost. Furthermore, it offers prediction quality that is either comparable to or sometimes surpasses that of other libraries. It is built on two key innovations: algorithmic and software-related.
The algorithmic innovation consists in a new strategy to speed up tree structure search in multioutput setup by approximating ("sketching") the scoring function used to find optimal splits. The approximation is achieved by reducing the dimensions of the gradient and Hessian matrices, while keeping the other boosting steps unchanged. This enables a significant speed-up for the main bottleneck in the boosting algorithm.
The software innovation is developed in a library that facilitates the efficient execution of complex algorithms (e.g., boosting) directly in Python while leveraging GPU capabilities. This means you can write straightforward Python code that approaches the efficiency of low-level optimized C code, thanks to the GPU utilization. It is worth to emphasize that despite "sketching" provides an approximation of the scoring function and thus one may expect some loss of prediction quality, we often observe an opposite effect - the approximation sometimes produce better predictions than the original function. 
This occurs because the simplified ("sketched") loss function appears to have better generalization capability for unseen data.

Another advantage of Py-Boost is its ease of use. The package can be installed by standard Python commands. The training process and parameter settings is very similar to other boosting packages such as XGBoost, LightGBM and CatBoost. Py-Boost served as our core model. We incorporated features from the Prot-T5 protein language model, combined with one-shot taxon features, and used Py-Boost to predict 4,500 targets in a single run. The training process for one fold took approximately 2 hours on a single GPU. Another variant of the Py-Boost model was obtained by the conditional probability modeling scheme described below.

\subsection{Protein language models}
Protein language models provide a new technique for protein sequence analysis which is borrowed from the modern developments in natural language processing. One of the ideas is that for each sequence protein sequence these models can generate an "embedding" i.e. a numerical vector. These vectors were designed to capture the essential properties of proteins, meaning that similar proteins should have corresponding similar vector representations. Training large protein language models require lots of computational power,
the top models were trained by Meta \cite{lin2022language} (ESM2-family) and by the 
B.Rost team \cite{elnaggar2021prottrans} (e.g. Prot-T5) using supercomputers.
The training set consists of hundreds millions of protein sequences. 
The length of the vectors does not depend on the length of the protein sequence, very short and very long protein sequences will be represented by vectors of the same length. The length of the vectors should ideally reflect the model's capacity—stronger models may have vector lengths around a thousand, while weaker models might have lengths around a hundred. However, this is not always achieved in practice, as overly large and complex models can easily overfit. These models are trained in semi-supervised manner, without any knowledge of any specific task, such as predicting Gene Ontology annotations. Therefore, to apply them to a specific predictive problem, we designed a custom model built on top of the embeddings. We used embeddings from Prot-T5 as features for our predictive models. 
The code to  generate the Prot-T5 embeddings can be found here:\newline
\href{https://www.kaggle.com/code/sergeifironov/t5embeds-calculation-only-few-samples}{https://www.kaggle.com/code/sergeifironov/t5embeds-calculation-only-few-samples}.

\begin{figure}[h]
\centering
\includegraphics[width=0.9\textwidth]{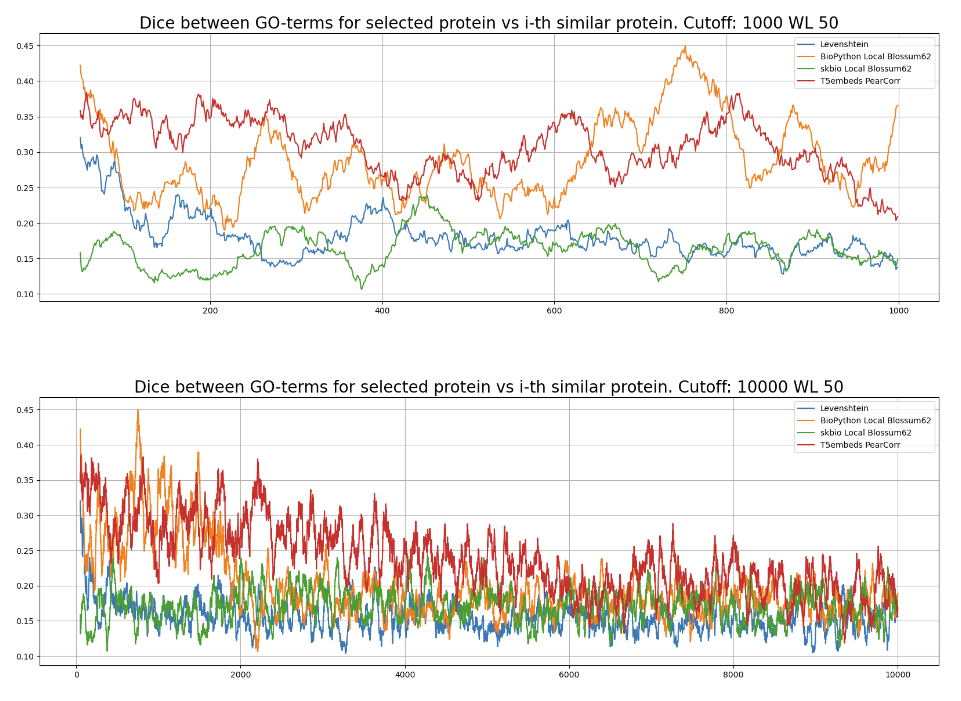}
\caption{Comparison of different structural similarities for proteins vs. functional similarity (dice-index for GO-terms). Y-axis reports  functional similarity between the selected protein (SRC human)
and the i-th structurally similar protein. Prot-T5 embedding shows superior similarity over other alignment methods.
Results are smoothed by moving average of length 50.  
}\label{compare_protein_similarities_for_SRC}
\end{figure}

To demonstrate efficiency of the protein language models in capturing functional similarities between proteins we performed the following analysis.
We have taken a set of proteins, including SRC human proto-oncogene tyrosine-protein kinase, and ordered all our training data proteins
by structural similarity, using several approaches, including Prot-T5 embeddings and other standard alignment methods.  For the i-th similar protein we compute functional similarity between SRC and that protein. Functional similarity is plotted  on the Y-axis for "i"-position on the X-axis. The functional similarity
is computed as dice-index between GO-terms of the two proteins. Results are smoothed by moving average of length 50. Smoothing allows to mitigate the noisy nature of the Gene Ontology annotations, where almost identical proteins might have quite different labels,
which makes plots too much oscillatory.
We observe that Prot-T5 embeddings show the most closest functional  similarity (see Figure \ref{compare_protein_similarities_for_SRC}). 
The line corresponding to Prot-T5 embedding is significantly above the other lines. This indicates that functional similarity aligns more closely with Prot-T5 similarity than with other types of similarity. 

The notebook to reproduce the analysis is available here:
\href{https://www.kaggle.com/code/alexandervc/cafa5-21-embed-beats-align-cases-src-p53}{https://www.kaggle.com/code/alexandervc/cafa5-21-embed-beats-align-cases-src-p53}.

\subsection{Conditional probability modeling}
We propose a specific scheme to work with targets which belong to a hierarchical structure like Ontology which we call "Conditional probability modelling scheme". This is not specific to protein function prediction task but can be applied to any task  where targets belong to a directed acyclic graph. It allows to construct from one machine learning model another modified one which is quite diverse from the original and ensembling these two models, achieving better prediction performance. 

During CAFA5, we applied this method to our Py-Boost and LogReg models, resulting in four models instead of two for building the ensemble. To apply the method one needs some modeling scheme for the task where targets belong to a directed acyclic graph. The idea is to modify the targets taking the graph structure into account, and then train the model on the new targets. To get the performance boost one ensembles the new model and the original one. Our focus is on the case of binary targets, though other target schemas can also be explored.
The method consists of the two steps:

\begin{itemize}
\item The first step is made during the training part: 
certain pairs of samples (proteins) and targets (GO-terms) are just excluded from the loss function. 
We exclude those pairs for which all GO-parents nodes have zero values.
(Technically we can just assign NA values for these pairs and loss function will ignore them).
Mathematically that means that we compute conditional probabilities under the condition that parent node is non-zero. 

\item The second step is made after the model is trained as a post-processing step.
We recompute the probabilities recursively according to the formula:
$P_{modified}(Node)= P_{model}(Node) ( 1 - \prod_{P:Parent~nodes} ( 1- P_{modified}(P) ) $. That is recursive process starting from the root node. The sense of the formula is the following: if parent probabilities are low, we need to lower the probabilities of the child node. 
The advantage of that approach is that it takes the hierarchical tree
structure of the targets into account. That is: GO-ontology use is built-in in that formula. So it might not be surprising that the performance 
for such a modification is better since it uses the graph structure
explicitly. But the main use is to ensemble 
the modified model with the original one, which boosts
the performance significantly.

\end{itemize}

\subsection{Stacking with the graph neural network.}
We developed certain modification of the Graph Neural Networks (GCN) to ensemble predictions of all our models as well as  external data. This approach can also be interpreted as a form of stacking.

Stacking is the machine learning technique to 
aggregate predictions of several models using a new 
model. Predictions of the basic models are used
as features for the new model. 
Thus, it represents a type of ensembling technique. However, ensembling typically refers to a linear combination of predictions from base models, whereas the term "stacking" is generally used when aggregation is carried out by a more sophisticated model.

We developed an original technique
of stacking based on graph neural network ideas.
That serves two purposes. Firstly, stacking itself, involving the aggregation of predictions from multiple base models (as well as the incorporation of so-called electronic annotations). Secondly, the explicit utilization of the ontology graph structure. Both purposes serve for improving the model performance,
and it was quite effective - overall improvement
was from 0.59 to 0.62 (see section "ablation study").

Reminder on classical GCNs and the difference with our approach. 
and some target variables at each node of the graph -  GCN performs aggregations of the features 
from neighbor nodes, followed by the non-linearity like ReLu. 
The idea behind classical GCNs is as follows: given vector features and target variables associated with each node in a graph, a GCN aggregates the features from neighboring nodes and applies a non-linear activation function, such as ReLU, to the result.
Trainable parameters
typically control the aggregation process. 
Loss function is defined as sum over nodes measuring difference between GCN output for each node
and target values. 
In our case: graph is Gene Ontology graph, features - are mainly made out predictions
of the basic models  and targets are exactly GO labels. 
But there is some difference between ours setup and classical GCN,
each protein defines its own features and its own targets for each node of the graph.
From a classical perspective, one might interpret our approach as having many GCNs, each parameterized by a specific protein. However, this interpretation is not entirely accurate. Our loss function involves a double summation over both proteins and nodes, and the trainable parameters are shared across all proteins, rather than being unique to each.
Thus, it is not a collection of several independent GCNs but rather a more sophisticated and interconnected construction.

In the classical setup, the train-test split is performed on the nodes of the graph, whereas in our approach, the train-test split is based on proteins. Specifically, if a protein is in the training set, the entire graph associated with that protein, along with all its labels, is included in the training set.



\begin{figure}[h]
\centering
\includegraphics[width=0.9\textwidth]{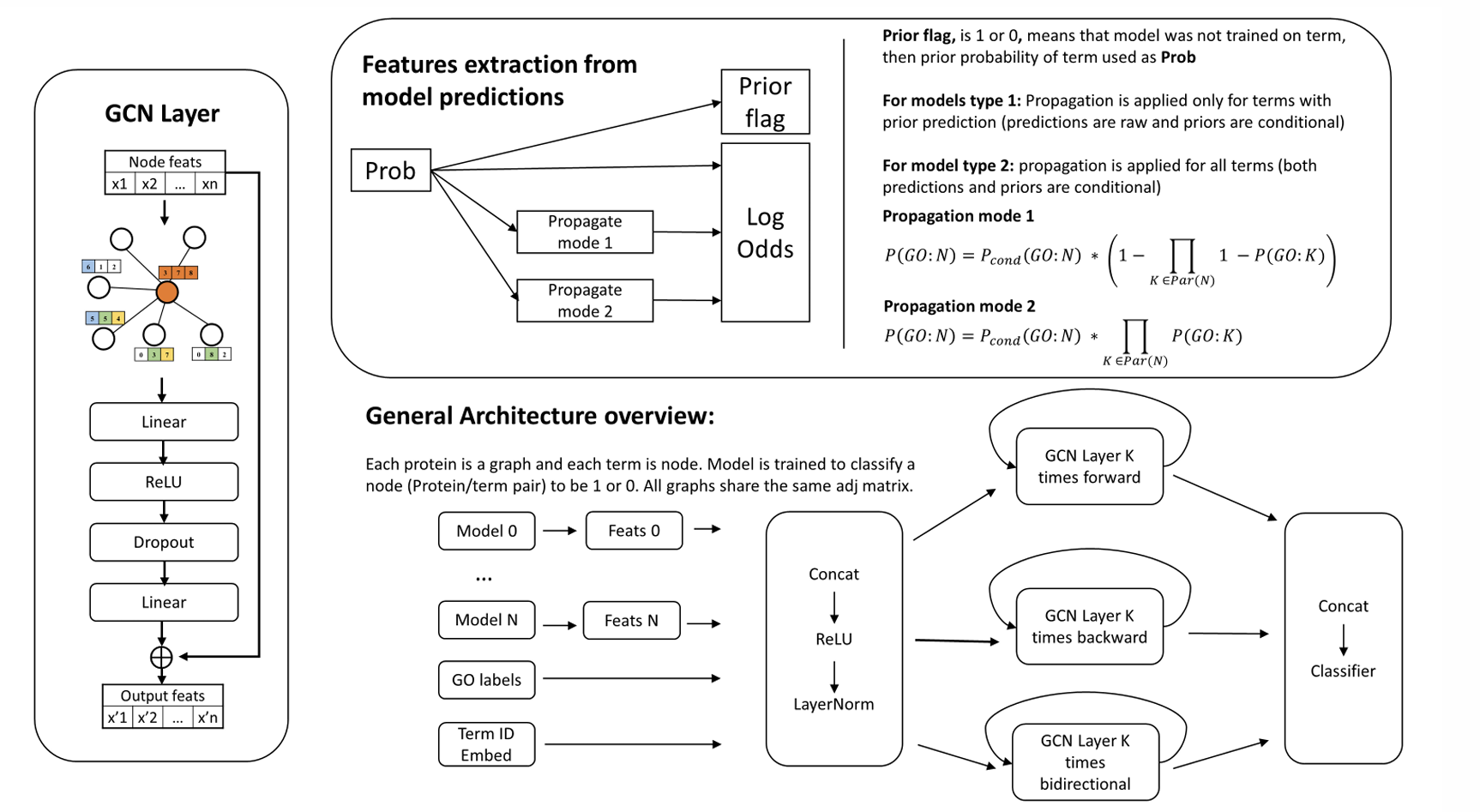}
\caption{Stacking with GCN (Graph Neural Network).
Model predictions are multiplexed to get 4 features from one prediction. That is done by two ways of propogations, prior flag and prediction itself (its logit). GCN Layer consists of Linear, ReLu, Dropout, Linear layers. Model features are concantenaded with "electoronic annotations" from the Uniprot, and trainable embedding. GCN process layers corresponding to ontology edges which are  considered in three ways: as directed, as directed backwards and bidirectional }\label{stacking_CAFA}
\end{figure}

Detailed description. To give details 
we need to describe features and architecture of GCN. 
( The structure of GCN is illustrated by figure \ref{stacking_CAFA}. )

GCN features - detailed description.
In total there are 29 features, coming from three sources:
20 features are made out of 5  our basic models predictions ;
1 feature is  the external data downloaded from Uniprot - 
"electronic annotations" and 8 features  trainable
embedding vector for each node of the Gene Ontology. 
Each model prediction is multiplexed into 4 features.
One of them is just logit of the prediction. 
The other two are made by the two different ways of modifications of predictions using graph structure
(called "propagation").
The first way of "propagation" is defined by the formula:
$P_{modified}(Node)= P_{model}(Node) ( 1 - \prod_{P:Parent~nodes} ( 1- P_{modified}(P) ) $, and the second:
$P_{modified}(Node)= P_{model}(Node) ( \prod_{P:Parent~nodes}  P_{modified}(P) $.
Using these formulas we recursively modify the predictions - starting from the root node and going down to the terminal nodes. 
The first way is our main way
and will be discussed in the next section.
The second is less efficient, but GCN extracts useful
information even from it. The sense of the both formulas is quite simple - if
parent probabilities are low we need to lower the probabilities of the child node.
We trained models on different numbers of targets,
so to have a common basis - we put prior - if 
that target was absent in training for the current model.
And the last feature is very simple 
it is binary flag 0,1 - indicating 
if prior has been used or just the direct model prediction.
So 4 features out of 5 models - give 20 features.
One more additional feature is the "electronic" GO annotation
downloaded from the Uniprot. 
And 8 more are just some trainable vector specific for each node.
I.e. an embedding of the GO-term.
So in total there are 29 features = 5*4 + 1 + 8.

GCN architecture.
The structure of our  basic layer is the following:
the features from the neighbor nodes of a given node
are averaged , and concatenated with maximum of these features.
That is followed by the standard layers: linear, ReLu, Dropout and again Linear.
The output is a vector ("embedding", aggregating information from neighbor features). 
Yet another detail: we have two layers with independently trained weights - the first one
takes 29 features described above and transform them into embedding.
The internal layers which are repeated 8 times - have the same dimension of the input embedding
and output embedding. The output of the first iteration is input for the second and so on.
We have three sorts of such layers - those for directed graph,
directed in opposite direction and undirected. 
Computation of GCN layers is repeated 8 times, also residual connections are used.
Outputs of all these layers are concatenated and 
passed to the classifier. 
The loss function - is standard BCE - which compares the output of the classifier and 
true targets. Pay attention again: the loss is obtained by summation over all proteins and nodes of the graph,
in contrast to classical GCN where summation is only over nodes. 

In conclusion, we developed a suitable modification of the GCN
framework, which is an innovative way to combine predictions of multiple models into ensembled one and moreover profits from utilizing the Gene Ontology graph structure.
It is proved to be quite effective - boosting the score significantly. 

\subsection{Postprocessing }
The final step of ProtBoost model consists in postprocessing results. 
It is implemented on top of the GCN stacking output predictions. 
The idea behind is quite simple. It is expected that probability 
of the parent node is greater or equal than probability of the child node.
Moreover it is explicitly implied by CAFA evaluation procedure,
where the first step of the processing, called "propagation",
is to update submitted predictions to make parents nodes greater or equal than child nodes. CAFA propagation goes from terminal nodes to the root.

We use two variants of the "propagation" and final results is their average.
The first one (denoted $P_{min}$ is given by the formula:
$P_{min}(N) = Min(P_{min}(k):k \in Parent(N))*0,7 + P(N)*0.3 $
and computed recursively traversing the graph from root to terminal nodes.
The second way (denoted $P_{max}$) is :
$P_{max}(N) = Min(P_{max}(m):m \in Child(N))*0,7 + P(N)*0.3 $
and computed recursively traversing the graph from terminal nodes to the root node.
And the final result is their average: $P_{postprocessing}(N) = \frac{P_{max}(N)+P_{min}(N) } {2}$

This final postprocessing step ensures a better alignment with the expected pattern of output probabilities, wherein parent nodes are assigned higher probabilities than their child nodes. 

\begin{figure}[h]
\centering
\includegraphics[width=0.9\textwidth]{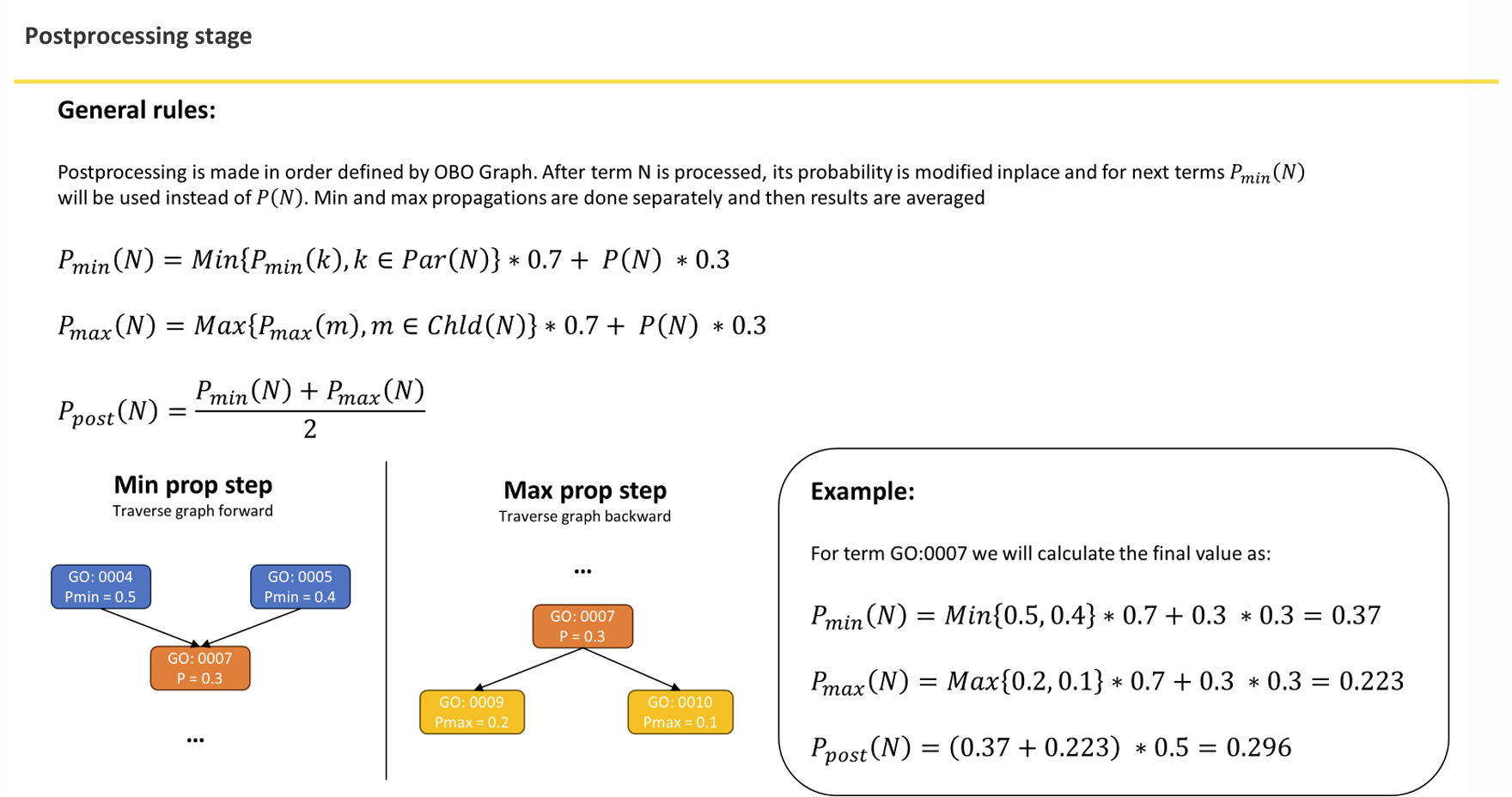}
\caption{Postprocessing step. The idea behind posprocessing is to leverage the property the probabilities of the parent nodes should be greater or equal than child nodes. That can be done traversing the graph from root to terminal nodes or from terminal to root nodes. We combine both way and average them. 
The process is done in the soft way - just taking the weighted average with coefficients 0.7 and 0.3 of the parent and child nodes.  }\label{fig1}
\end{figure}

\subsection{Machine learning models}

First, we developed\cite{baselineCAFA5} a baseline model using Prot-T5 embeddings and the Ridge model on top. As the next steps we tried to substitute Ridge by more powerful models. 
Surprisingly many of them fail to show good performance:
direct application of the sklearn logistic regression fails to converge and we used GPU based version
with control of iteration number in our final solution. Gradient boosting like LightGBM fails due to time limits to work on multi-target tasks, but even restricted to 500 targets and trained for more than 24h in parallel on 10 virtual machines, showed results worse than Ridge regression. SVM, Kernel Ridge, Random Forest - are too slow on multi-target task, but even restricted to 100 targets showed worse performance than Ridge. 
This contrasts with our primary method, Py-Boost, which demonstrated effective performance. Finally, to complement the Py-Boost model,  we developed  neural network models, with rather simple architecture of 2-layer perceptron with droupouts and batchnorms.

Regarding features, we observed that the Prot-T5 language model performs best; however, its performance can be further enhanced by concatenating it with taxon (one-hot) and ESM2 embeddings.

\subsection{Timing}

The model was trained on 142,246 samples and evaluated on 141,865 samples. Embeddings were precomputed; embedding inference takes approximately 3 hours per run. All times are reported for a single V100 32GB GPU.

Overall.
Training: minimal resources (single GPU) ~ 4.5 days;  optimal (3 GPUs) ~ 2 days;
fastest (all parallel, 7 GPUs) ~ 1.4 days;   inference only (Single GPU) ~ 3 hours. 

For each component. (single run, 5 fold CV):
py-boost:15 hours;  LogReg:2-10 hours; NeuralNet:1 hour. Stacking train: 2-13 hours.
Stacking inference (4 times): 1.5 hours– 20 min;  Postprocessing: 5 minutes.

\subsection{Gene Ontology}
The Gene Ontology (GO) is one of the key databases describing gene functions and properties across all species\cite{ashburner2000gene}. CAFA5 was based on the release 2023-01-01, which contains around 40 thousands different gene annotation terms.
It is divided into three separate ontologies:  Molecular Functions (MF), which describes the activities of a gene product at the molecular level, such as binding or catalysis; Cellular Component (CC), which describes  the locations, relative to cellular structures, where MFs are performed; Biological Process (BP), a “biological program” comprising molecular activities acting in concert to achieve a particular outcome; this program can be at the cellular level or at the organism level of multicellular organisms.
The GO database is organized as a directed acyclic graph, where each term is connected to one or more other terms within the same domain. This vocabulary is species-neutral, encompassing terms relevant to both prokaryotes and eukaryotes, as well as to single-celled and multicellular organisms.

More recent versions of GO allows connections between terms from different sub-ontologies, i.e. molecular function terms can be connected to biological process terms, 
but the version used in CAFA5 followed earlier policy where such connections were forbidden.
Evaluation in CAFA challenges is based on suitable modification F1-metric, however before the evaluation,
the so-called propagation process is computed. That means predictions of "parent" nodes are made 
greater or equal than predictions of the child nodes. A crucial step in many solutions, including our model, involved specialized post-processing that achieves this specific feature.

The training set contains 31466 different GO terms (BPO:21285, CCO:2957,MFO:7224). Very few methods \cite{kulmanov2024protein} allow non-trivial zero-shot predictions for such terms, including those entirely absent from the training set. Our model enables to predict 13000 top frequent terms. 

\subsection{Datasets}\label{ss_data}

CAFA5 organizers provided the training set with 142,246 protein sequences and Gene Ontology annotations borrowed from the UniProtKB \cite{uniprot2023uniprot} release 2022-11-17. A second dataset, called the “super-test” dataset, consisting of 141,865 sequences, was provided to participants to generate their predictions. The training proteins come from more than 3000 different species, while the super-test only from 90 species. An unknown subset of these protein sequences constituted the actual test set(s) for evaluation, meaning participants were asked to submit predictions for all 141,865 sequences, but the actual assessment was conducted on a small undisclosed portion.
Furthermore, the evaluation was divided into two phases. The first phase, the preliminary or "public" leaderboard, was accessible to participants immediately during the challenge. It was  allowed to make up to five submissions per day. The submission results were scored on the "public" part of test set
and scores were shown to participants immediately. The experimental labels for these proteins
were unknown from all participants and known only to the organizers. The second phase of the evaluation, which included results from new experiments conducted by the organizers, became accessible only four months after the challenge ended.
So during the challenge the experimental labels for these proteins were completely unknown to everyone \textit{including} the organizers, organizers performed experiments after the end of the computational part of the challenge. This constituted the final evaluation.

Explanatory data analysis of the training and testing data was performed and available here: \href{https://www.kaggle.com/code/alexandervc/cafa5-towards-eda}
{https://www.kaggle.com/code/alexandervc/cafa5-towards-eda}.

\subsection{Cross-validation}

There are several possible cross-validation strategies which  were used by CAFA participants. One option (used e.g. by Top3 team \cite{top3CAFA5})
The approach relies on the concept of a temporal split, wherein the training set includes proteins annotated before a specific date, while the validation set comprises those annotated afterward. 
Another option is to build the training and testing dataset by protein similarity, that is to put in the validation set proteins which are the most dissimilar to the  training set. This scheme leverages the ability of the model to generalize  to the unseen proteins. 
The scheme we adopted primarily involved random train-test splits into five folds. However, we also incorporated the concept of time-split validation, albeit on a limited scale. Specifically, we selected a small subset of proteins that were annotated relatively recently and designated them as our final small test holdout fold, utilized during the stacking phase.

\subsection{Feature engineering}
The main features we used were based on the protein language models discussed above.
We also employed taxons information for most frequent taxons with the
one-hot encoding. These taxons features showed quite a good
improvement for all our models and validation schemes. 
This is expected since some GO terms might be specific for certain types of organisms.
Furtheromore, we parsed GO information from Uniprot database, available at \href{http://ftp.ebi.ac.uk/pub/databases/GO/goa/old/UNIPROT/}{http://ftp.ebi.ac.uk/pub/databases/GO/goa/old/UNIPROT/},
to get  the most actual GO labeling. We used these electronic
 labels as features at stacking  to predict  experimental labels.
 
\subsection{Evaluation metric}\label{ss_eval_m}
The evaluation metric in the CAFA challenge is a modified F1-metric.
Scores are computed for each sub-ontology separately and then averaged.
GO terms are incorporated into the metric with specific weights assigned by the organizers. These weights are designed to emphasize terms that are rarer and, consequently, more challenging to predict.

Also continuous predictions were binarized to 0 and 1 before computing F1-metric, which was done by checking all the possible
binarization thresholds from 0 to 1 with step 0.01 and choosing the maximum value among all obtained weighted F1. In addition, before the metric computation the so-called propagation process on predictions
is executed. It ensures that parent GO-nodes have predictions greater or equal than child nodes. That propagation process proceeds from terminal nodes to the root nodes.

We improved this metric computation procedure to make it less time consuming. This implementation, available in the file "metric.py" in our github repository, contains both algorithmic improvements as well as a GPU version significantly faster. Our implementation was adopted by the organizers. 

Another function we improved was the function "\verb|compute_metrics_|"
where we can use sparse matrix multiplication for 
computations of variable intersection. 
By utilizing GPU acceleration, we achieved a more than tenfold speed-up in CAFA metric computation, reducing the runtime from approximately one hour to just a few minutes.

\section{Data availability}
Main data underlying this work
is the official data of the CAFA5 challenge freely available at : \newline
\href{https://www.kaggle.com/competitions/cafa-5-protein-function-prediction/data}{https://www.kaggle.com/competitions/cafa-5-protein-function-prediction/data}. 

We computed Prot-T5 embeddings and freely shared them here:
\href{https://www.kaggle.com/datasets/sergeifironov/t5embeds}{https://www.kaggle.com/datasets/sergeifironov/t5embeds}

EMS2 embeddings can be found freely at: \href{https://www.kaggle.com/competitions/cafa-5-protein-function-prediction/discussion/406168}{https://www.kaggle.com/competitions/cafa-5-protein-function-prediction/discussion/406168}

Additionally, we computed and shared Prot-T5 embeddings for more than 500 thousands
proteins (Uniprot manually curated part): \newline
\href{https://www.kaggle.com/competitions/cafa-5-protein-function-prediction/discussion/466703}{https://www.kaggle.com/competitions/cafa-5-protein-function-prediction/discussion/466703}, 
ESM2 (large - dimension more than 5000) and Ankh models: \newline
\href{https://www.kaggle.com/competitions/cafa-5-protein-function-prediction/discussion/462419}{https://www.kaggle.com/competitions/cafa-5-protein-function-prediction/discussion/462419}.

\section{Code availability}

The source code of this work is freely available at :\newline
\href{https://github.com/btbpanda/CAFA5-protein-function-prediction-2nd-place}{https://github.com/btbpanda/CAFA5-protein-function-prediction-2nd-place}

The code we generated for Prot-T5 embeddings is available at:\newline
\href{https://www.kaggle.com/code/sergeifironov/t5embeds-calculation-only-few-samples}{https://www.kaggle.com/code/sergeifironov/t5embeds-calculation-only-few-samples}.

The prototype code for our neural network is available at:\newline
\href{https://www.kaggle.com/code/alexandervc/pytorch-keras-etc-3-blend-cafa-metric-etc}{https://www.kaggle.com/code/alexandervc/pytorch-keras-etc-3-blend-cafa-metric-etc}

\bibliography{sn-bibliography}


\begin{thebibliography}{25}
\ifx \bisbn   \undefined \def \bisbn  #1{ISBN #1}\fi
\ifx \binits  \undefined \def \binits#1{#1}\fi
\ifx \bauthor  \undefined \def \bauthor#1{#1}\fi
\ifx \batitle  \undefined \def \batitle#1{#1}\fi
\ifx \bjtitle  \undefined \def \bjtitle#1{#1}\fi
\ifx \bvolume  \undefined \def \bvolume#1{\textbf{#1}}\fi
\ifx \byear  \undefined \def \byear#1{#1}\fi
\ifx \bissue  \undefined \def \bissue#1{#1}\fi
\ifx \bfpage  \undefined \def \bfpage#1{#1}\fi
\ifx \blpage  \undefined \def \blpage #1{#1}\fi
\ifx \burl  \undefined \def \burl#1{\textsf{#1}}\fi
\ifx \doiurl  \undefined \def \doiurl#1{\url{https://doi.org/#1}}\fi
\ifx \betal  \undefined \def \betal{\textit{et al.}}\fi
\ifx \binstitute  \undefined \def \binstitute#1{#1}\fi
\ifx \binstitutionaled  \undefined \def \binstitutionaled#1{#1}\fi
\ifx \bctitle  \undefined \def \bctitle#1{#1}\fi
\ifx \beditor  \undefined \def \beditor#1{#1}\fi
\ifx \bpublisher  \undefined \def \bpublisher#1{#1}\fi
\ifx \bbtitle  \undefined \def \bbtitle#1{#1}\fi
\ifx \bedition  \undefined \def \bedition#1{#1}\fi
\ifx \bseriesno  \undefined \def \bseriesno#1{#1}\fi
\ifx \blocation  \undefined \def \blocation#1{#1}\fi
\ifx \bsertitle  \undefined \def \bsertitle#1{#1}\fi
\ifx \bsnm \undefined \def \bsnm#1{#1}\fi
\ifx \bsuffix \undefined \def \bsuffix#1{#1}\fi
\ifx \bparticle \undefined \def \bparticle#1{#1}\fi
\ifx \barticle \undefined \def \barticle#1{#1}\fi
\bibcommenthead
\ifx \bconfdate \undefined \def \bconfdate #1{#1}\fi
\ifx \botherref \undefined \def \botherref #1{#1}\fi
\ifx \url \undefined \def \url#1{\textsf{#1}}\fi
\ifx \bchapter \undefined \def \bchapter#1{#1}\fi
\ifx \bbook \undefined \def \bbook#1{#1}\fi
\ifx \bcomment \undefined \def \bcomment#1{#1}\fi
\ifx \oauthor \undefined \def \oauthor#1{#1}\fi
\ifx \citeauthoryear \undefined \def \citeauthoryear#1{#1}\fi
\ifx \endbibitem  \undefined \def \endbibitem {}\fi
\ifx \bconflocation  \undefined \def \bconflocation#1{#1}\fi
\ifx \arxivurl  \undefined \def \arxivurl#1{\textsf{#1}}\fi
\csname PreBibitemsHook\endcsname

\bibitem[\protect\citeauthoryear{Ashburner et~al.}{2000}]{ashburner2000gene}
\begin{barticle}
\bauthor{\bsnm{Ashburner}, \binits{M.}},
\bauthor{\bsnm{Ball}, \binits{C.A.}},
\bauthor{\bsnm{Blake}, \binits{J.A.}},
\bauthor{\bsnm{Botstein}, \binits{D.}},
\bauthor{\bsnm{Butler}, \binits{H.}},
\bauthor{\bsnm{Cherry}, \binits{J.M.}},
\bauthor{\bsnm{Davis}, \binits{A.P.}},
\bauthor{\bsnm{Dolinski}, \binits{K.}},
\bauthor{\bsnm{Dwight}, \binits{S.S.}},
\bauthor{\bsnm{Eppig}, \binits{J.T.}}, \betal:
\batitle{Gene ontology: tool for the unification of biology}.
\bjtitle{Nature genetics}
\bvolume{25}(\bissue{1}),
\bfpage{25}--\blpage{29}
(\byear{2000})
\end{barticle}
\endbibitem

\bibitem[\protect\citeauthoryear{}{2023}]{uniprot2023uniprot}
\begin{botherref}
Uniprot: the universal protein knowledgebase in 2023.
Nucleic acids research
\textbf{51}(D1),
523--531
(2023)
\end{botherref}
\endbibitem

\bibitem[\protect\citeauthoryear{Lee et~al.}{2007}]{lee2007predicting}
\begin{barticle}
\bauthor{\bsnm{Lee}, \binits{D.}},
\bauthor{\bsnm{Redfern}, \binits{O.}},
\bauthor{\bsnm{Orengo}, \binits{C.}}:
\batitle{Predicting protein function from sequence and structure}.
\bjtitle{Nature reviews molecular cell biology}
\bvolume{8}(\bissue{12}),
\bfpage{995}--\blpage{1005}
(\byear{2007})
\end{barticle}
\endbibitem

\bibitem[\protect\citeauthoryear{Tiwari and Srivastava}{2014}]{tiwari2014survey}
\begin{barticle}
\bauthor{\bsnm{Tiwari}, \binits{A.K.}},
\bauthor{\bsnm{Srivastava}, \binits{R.}}:
\batitle{A survey of computational intelligence techniques in protein function prediction}.
\bjtitle{International journal of proteomics}
\bvolume{2014}(\bissue{1}),
\bfpage{845479}
(\byear{2014})
\end{barticle}
\endbibitem

\bibitem[\protect\citeauthoryear{Radivojac et~al.}{2013}]{radivojac2013large}
\begin{barticle}
\bauthor{\bsnm{Radivojac}, \binits{P.}},
\bauthor{\bsnm{Clark}, \binits{W.T.}},
\bauthor{\bsnm{Oron}, \binits{T.R.}},
\bauthor{\bsnm{Schnoes}, \binits{A.M.}},
\bauthor{\bsnm{Wittkop}, \binits{T.}},
\bauthor{\bsnm{Sokolov}, \binits{A.}},
\bauthor{\bsnm{Graim}, \binits{K.}},
\bauthor{\bsnm{Funk}, \binits{C.}},
\bauthor{\bsnm{Verspoor}, \binits{K.}},
\bauthor{\bsnm{Ben-Hur}, \binits{A.}}, \betal:
\batitle{A large-scale evaluation of computational protein function prediction}.
\bjtitle{Nature methods}
\bvolume{10}(\bissue{3}),
\bfpage{221}--\blpage{227}
(\byear{2013})
\end{barticle}
\endbibitem

\bibitem[\protect\citeauthoryear{Jiang et~al.}{2016}]{jiang2016expanded}
\begin{barticle}
\bauthor{\bsnm{Jiang}, \binits{Y.}},
\bauthor{\bsnm{Oron}, \binits{T.R.}},
\bauthor{\bsnm{Clark}, \binits{W.T.}},
\bauthor{\bsnm{Bankapur}, \binits{A.R.}},
\bauthor{\bsnm{D’Andrea}, \binits{D.}},
\bauthor{\bsnm{Lepore}, \binits{R.}},
\bauthor{\bsnm{Funk}, \binits{C.S.}},
\bauthor{\bsnm{Kahanda}, \binits{I.}},
\bauthor{\bsnm{Verspoor}, \binits{K.M.}},
\bauthor{\bsnm{Ben-Hur}, \binits{A.}}, \betal:
\batitle{An expanded evaluation of protein function prediction methods shows an improvement in accuracy}.
\bjtitle{Genome biology}
\bvolume{17},
\bfpage{1}--\blpage{19}
(\byear{2016})
\end{barticle}
\endbibitem

\bibitem[\protect\citeauthoryear{Zhou et~al.}{2019}]{zhou2019cafa}
\begin{barticle}
\bauthor{\bsnm{Zhou}, \binits{N.}},
\bauthor{\bsnm{Jiang}, \binits{Y.}},
\bauthor{\bsnm{Bergquist}, \binits{T.R.}},
\bauthor{\bsnm{Lee}, \binits{A.J.}},
\bauthor{\bsnm{Kacsoh}, \binits{B.Z.}},
\bauthor{\bsnm{Crocker}, \binits{A.W.}},
\bauthor{\bsnm{Lewis}, \binits{K.A.}},
\bauthor{\bsnm{Georghiou}, \binits{G.}},
\bauthor{\bsnm{Nguyen}, \binits{H.N.}},
\bauthor{\bsnm{Hamid}, \binits{M.N.}}, \betal:
\batitle{The cafa challenge reports improved protein function prediction and new functional annotations for hundreds of genes through experimental screens}.
\bjtitle{Genome biology}
\bvolume{20},
\bfpage{1}--\blpage{23}
(\byear{2019})
\end{barticle}
\endbibitem

\bibitem[\protect\citeauthoryear{Iosipoi and Vakhrushev}{2022}]{vakhrushev2022pyboost}
\begin{barticle}
\bauthor{\bsnm{Iosipoi}, \binits{L.}},
\bauthor{\bsnm{Vakhrushev}, \binits{A.}}:
\batitle{Sketchboost: Fast gradient boosted decision tree for multioutput problems}.
\bjtitle{Advances in Neural Information Processing Systems}
\bvolume{35},
\bfpage{25422}--\blpage{25435}
(\byear{2022}).
\bcomment{\url{https://youtu.be/5xRxuDh_cGk}}
\end{barticle}
\endbibitem

\bibitem[\protect\citeauthoryear{Altschul et~al.}{1990}]{altschul1990basic}
\begin{barticle}
\bauthor{\bsnm{Altschul}, \binits{S.F.}},
\bauthor{\bsnm{Gish}, \binits{W.}},
\bauthor{\bsnm{Miller}, \binits{W.}},
\bauthor{\bsnm{Myers}, \binits{E.W.}},
\bauthor{\bsnm{Lipman}, \binits{D.J.}}:
\batitle{Basic local alignment search tool}.
\bjtitle{Journal of molecular biology}
\bvolume{215}(\bissue{3}),
\bfpage{403}--\blpage{410}
(\byear{1990})
\end{barticle}
\endbibitem

\bibitem[\protect\citeauthoryear{Wang et~al.}{2023}]{top1CAFA5}
\begin{botherref}
\oauthor{\bsnm{Wang}, \binits{S.}},
\oauthor{\bsnm{Zhai}, \binits{W.}},
\oauthor{\bsnm{Liu}, \binits{W.}},
\oauthor{\bsnm{Huang}, \binits{T.}},
\oauthor{\bsnm{Yan}, \binits{H.}}:
1st Place Solution for the CAFA5.
\url{https://www.kaggle.com/competitions/cafa-5-protein-function-prediction/discussion/466917}
(2023)
\end{botherref}
\endbibitem

\bibitem[\protect\citeauthoryear{Chua et~al.}{2024}]{chua2024protgoat}
\begin{botherref}
\oauthor{\bsnm{Chua}, \binits{Z.M.}},
\oauthor{\bsnm{Rajesh}, \binits{A.}},
\oauthor{\bsnm{Sinha}, \binits{S.}},
\oauthor{\bsnm{Adams}, \binits{P.D.}}:
Protgoat: Improved automated protein function predictions using protein language models.
bioRxiv,
2024--04
(2024).
\url{https://www.biorxiv.org/content/10.1101/2024.04.01.587572v1}
\end{botherref}
\endbibitem

\bibitem[\protect\citeauthoryear{Tito}{2023}]{top3CAFA5}
\begin{botherref}
\oauthor{\bsnm{Tito}}:
3rd Place Solution for the CAFA 5 Protein Function Prediction.
\url{https://www.kaggle.com/competitions/cafa-5-protein-function-prediction/discussion/464437}
(2023)
\end{botherref}
\endbibitem

\bibitem[\protect\citeauthoryear{Zhang and Freddolino}{2023}]{top5CAFA5}
\begin{botherref}
\oauthor{\bsnm{Zhang}, \binits{C.}},
\oauthor{\bsnm{Freddolino}, \binits{L.}}:
5th Place Solution for the CAFA 5 Protein Function Prediction Challenge.
\url{https://www.kaggle.com/competitions/cafa-5-protein-function-prediction/discussion/463009}
(2023)
\end{botherref}
\endbibitem

\bibitem[\protect\citeauthoryear{Liu}{2023}]{top6CAFA5}
\begin{botherref}
\oauthor{\bsnm{Liu}, \binits{Q.}}:
6th Place Solution for the CAFA 5 Protein Function Prediction Challenge.
\url{https://www.kaggle.com/competitions/cafa-5-protein-function-prediction/discussion/466971}
(2023)
\end{botherref}
\endbibitem

\bibitem[\protect\citeauthoryear{Chervov}{2023}]{top14CAFA5}
\begin{botherref}
\oauthor{\bsnm{Chervov}, \binits{A.}}:
3 remarkable ideas from Zoltan's notebook "CombineEmbeddings".
\url{https://www.kaggle.com/competitions/cafa-5-protein-function-prediction/discussion/431491}
(2023)
\end{botherref}
\endbibitem

\bibitem[\protect\citeauthoryear{Tinti}{2023}]{MTmergedatasets}
\begin{botherref}
\oauthor{\bsnm{Tinti}, \binits{M.}}:
Merge datasets.
\url{https://www.kaggle.com/code/mtinti/merge-datasets}
(2023)
\end{botherref}
\endbibitem

\bibitem[\protect\citeauthoryear{Kulmanov et~al.}{2024}]{kulmanov2024protein}
\begin{barticle}
\bauthor{\bsnm{Kulmanov}, \binits{M.}},
\bauthor{\bsnm{Guzm{\'a}n-Vega}, \binits{F.J.}},
\bauthor{\bsnm{Duek~Roggli}, \binits{P.}},
\bauthor{\bsnm{Lane}, \binits{L.}},
\bauthor{\bsnm{Arold}, \binits{S.T.}},
\bauthor{\bsnm{Hoehndorf}, \binits{R.}}:
\batitle{Protein function prediction as approximate semantic entailment}.
\bjtitle{Nature Machine Intelligence}
\bvolume{6}(\bissue{2}),
\bfpage{220}--\blpage{228}
(\byear{2024}).
\bcomment{\url{https://youtu.be/vhnD4SR8cWI}}
\end{barticle}
\endbibitem

\bibitem[\protect\citeauthoryear{Elnaggar et~al.}{2021}]{elnaggar2021prottrans}
\begin{barticle}
\bauthor{\bsnm{Elnaggar}, \binits{A.}},
\bauthor{\bsnm{Heinzinger}, \binits{M.}},
\bauthor{\bsnm{Dallago}, \binits{C.}},
\bauthor{\bsnm{Rehawi}, \binits{G.}},
\bauthor{\bsnm{Wang}, \binits{Y.}},
\bauthor{\bsnm{Jones}, \binits{L.}},
\bauthor{\bsnm{Gibbs}, \binits{T.}},
\bauthor{\bsnm{Feher}, \binits{T.}},
\bauthor{\bsnm{Angerer}, \binits{C.}},
\bauthor{\bsnm{Steinegger}, \binits{M.}}, \betal:
\batitle{Prottrans: Toward understanding the language of life through self-supervised learning}.
\bjtitle{IEEE transactions on pattern analysis and machine intelligence}
\bvolume{44}(\bissue{10}),
\bfpage{7112}--\blpage{7127}
(\byear{2021})
\end{barticle}
\endbibitem

\bibitem[\protect\citeauthoryear{Lin et~al.}{2022}]{lin2022language}
\begin{barticle}
\bauthor{\bsnm{Lin}, \binits{Z.}},
\bauthor{\bsnm{Akin}, \binits{H.}},
\bauthor{\bsnm{Rao}, \binits{R.}},
\bauthor{\bsnm{Hie}, \binits{B.}},
\bauthor{\bsnm{Zhu}, \binits{Z.}},
\bauthor{\bsnm{Lu}, \binits{W.}},
\bauthor{\bsnm{Santos~Costa}, \binits{A.}},
\bauthor{\bsnm{Fazel-Zarandi}, \binits{M.}},
\bauthor{\bsnm{Sercu}, \binits{T.}},
\bauthor{\bsnm{Candido}, \binits{S.}}, \betal:
\batitle{Language models of protein sequences at the scale of evolution enable accurate structure prediction}.
\bjtitle{BioRxiv}
\bvolume{2022},
\bfpage{500902}
(\byear{2022})
\end{barticle}
\endbibitem

\bibitem[\protect\citeauthoryear{Luukkonen et~al.}{2023}]{luukkonen2023large}
\begin{barticle}
\bauthor{\bsnm{Luukkonen}, \binits{S.}},
\bauthor{\bsnm{Meijer}, \binits{E.}},
\bauthor{\bsnm{Tricarico}, \binits{G.A.}},
\bauthor{\bsnm{Hofmans}, \binits{J.}},
\bauthor{\bsnm{Stouten}, \binits{P.F.}},
\bauthor{\bsnm{Westen}, \binits{G.J.}},
\bauthor{\bsnm{Lenselink}, \binits{E.B.}}:
\batitle{Large-scale modeling of sparse protein kinase activity data}.
\bjtitle{Journal of Chemical Information and Modeling}
\bvolume{63}(\bissue{12}),
\bfpage{3688}--\blpage{3696}
(\byear{2023})
\end{barticle}
\endbibitem

\bibitem[\protect\citeauthoryear{Jumper et~al.}{2021}]{jumper2021highly}
\begin{barticle}
\bauthor{\bsnm{Jumper}, \binits{J.}},
\bauthor{\bsnm{Evans}, \binits{R.}},
\bauthor{\bsnm{Pritzel}, \binits{A.}},
\bauthor{\bsnm{Green}, \binits{T.}},
\bauthor{\bsnm{Figurnov}, \binits{M.}},
\bauthor{\bsnm{Ronneberger}, \binits{O.}},
\bauthor{\bsnm{Tunyasuvunakool}, \binits{K.}},
\bauthor{\bsnm{Bates}, \binits{R.}},
\bauthor{\bsnm{{\v{Z}}{\'\i}dek}, \binits{A.}},
\bauthor{\bsnm{Potapenko}, \binits{A.}}, \betal:
\batitle{Highly accurate protein structure prediction with alphafold}.
\bjtitle{nature}
\bvolume{596}(\bissue{7873}),
\bfpage{583}--\blpage{589}
(\byear{2021})
\end{barticle}
\endbibitem

\bibitem[\protect\citeauthoryear{Grant et~al.}{2013}]{grant2013identification}
\begin{barticle}
\bauthor{\bsnm{Grant}, \binits{G.D.}},
\bauthor{\bsnm{Brooks~3rd}, \binits{L.}},
\bauthor{\bsnm{Zhang}, \binits{X.}},
\bauthor{\bsnm{Mahoney}, \binits{J.M.}},
\bauthor{\bsnm{Martyanov}, \binits{V.}},
\bauthor{\bsnm{Wood}, \binits{T.A.}},
\bauthor{\bsnm{Sherlock}, \binits{G.}},
\bauthor{\bsnm{Cheng}, \binits{C.}},
\bauthor{\bsnm{Whitfield}, \binits{M.L.}}:
\batitle{Identification of cell cycle--regulated genes periodically expressed in u2os cells and their regulation by foxm1 and e2f transcription factors}.
\bjtitle{Molecular biology of the cell}
\bvolume{24}(\bissue{23}),
\bfpage{3634}--\blpage{3650}
(\byear{2013})
\end{barticle}
\endbibitem

\bibitem[\protect\citeauthoryear{Giotti et~al.}{2019}]{giotti2019assembly}
\begin{barticle}
\bauthor{\bsnm{Giotti}, \binits{B.}},
\bauthor{\bsnm{Chen}, \binits{S.-H.}},
\bauthor{\bsnm{Barnett}, \binits{M.W.}},
\bauthor{\bsnm{Regan}, \binits{T.}},
\bauthor{\bsnm{Ly}, \binits{T.}},
\bauthor{\bsnm{Wiemann}, \binits{S.}},
\bauthor{\bsnm{Hume}, \binits{D.A.}},
\bauthor{\bsnm{Freeman}, \binits{T.C.}}:
\batitle{Assembly of a parts list of the human mitotic cell cycle machinery}.
\bjtitle{Journal of molecular cell biology}
\bvolume{11}(\bissue{8}),
\bfpage{703}--\blpage{718}
(\byear{2019})
\end{barticle}
\endbibitem

\bibitem[\protect\citeauthoryear{Chervov and Zinovyev}{2022}]{chervov2022computational}
\begin{botherref}
\oauthor{\bsnm{Chervov}, \binits{A.}},
\oauthor{\bsnm{Zinovyev}, \binits{A.}}:
Computational challenges of cell cycle analysis using single cell transcriptomics.
arXiv preprint arXiv:2208.05229
(2022)
\end{botherref}
\endbibitem

\bibitem[\protect\citeauthoryear{Chervov}{2023}]{baselineCAFA5}
\begin{botherref}
\oauthor{\bsnm{Chervov}, \binits{A.}}:
Baseline MultiLabel to MultiTarget(Binary).
\url{https://www.kaggle.com/code/alexandervc/baseline-multilabel-to-multitarget-binary}
(2023)
\end{botherref}
\endbibitem

\end{thebibliography}

\section*{Acknowledgments}

With financial support from ITMO Cancer of Aviesan within the framework of the 2021- 2030 Cancer Control Strategy, on funds administered by Inserm.
We are thankful to CAFA5 organizers and the Kaggle team to create such a great event.
We are also thankful to Emmanuel Barillot, Andrei Zinovyev, Loïc Chadoutaud, Olga Kalinina, Arthur Zalevsky, 
Yevgeniy Antipin, Liza Geraseva, Antonina Dolgorukova
for stimulating discussions and comments on the manuscript. 






\section*{Supplementary Material}

\begin{figure}[h]
\centering
\includegraphics[width=0.9\textwidth]{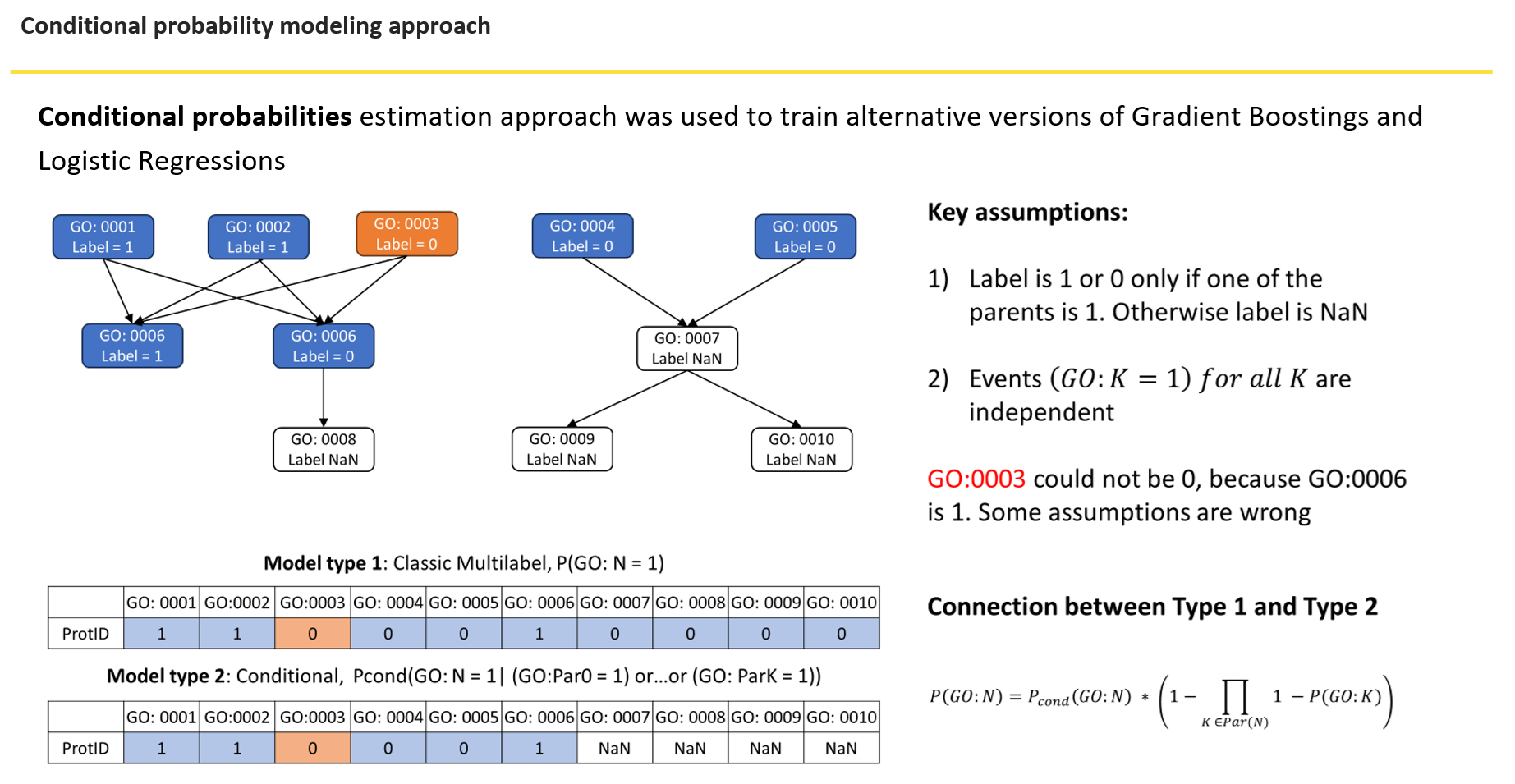}
\caption{Conditional probability modeling scheme (CondProbMod). Model loss includes  only those pairs of targets and samples  for which at least one parent node label is non-zero (e.g. GO:0008,GO:0007,GO:0009,GO:0010 are marked by "NaN" i.e. excluded from the loss function). Thus model itself predicts \textit{conditional} probabilities under the condition of existing a non-zero parent node. The postprocessing scheme computes actual predictions from conditional predicted probabilities
by the propagation along the ontology graph. Propagation here goes from the terminal to root nodes. }\label{fig1}
\end{figure}

\end{document}